\documentstyle[multicol,prl,aps]{revtex}
\draft
\begin{document}

\title
{
New mechanisms of droplet coarsening 
in phase-separating fluid mixtures
}
\author
{ 
Hajime Tanaka
}
\address
{
Institute of Industrial Science, University of Tokyo, Minato-ku, 
Tokyo 106, Japan. 
}
\date
{
Received 21 October 1996
}

\maketitle

\begin{abstract}
We propose here a new mechanism of droplet coarsening in 
phase-separating fluid mixtures. 
In contrast to the conventional understanding that there are 
no interactions between droplets in the late stage of spinodal 
decomposition, we demonstrate the existence of interactions 
between droplets that is caused by the coupling between diffuse 
concentration change around droplets. 
We show the possibility that this mechanism 
plays an important role in droplet phase separation together with 
Brownian-coagulation mechanism. 
We also discuss the coupling between hydrodynamic and diffusion modes, 
namely, "collision-induced collision" phenomena. 
\end{abstract}


\begin{multicols}{2}

Droplet patterns are generally observed in the late stage 
of phase separation in binary mixtures of solids and liquids 
at off-symmetric compositions \cite{Gunton,Bray}. 
There are two well-known coarsening mechanisms 
for droplet phase separation: 
(i) evaporation-condensation mechanism \cite{Gunton,LS} 
and (ii) Brownian-coagulation mechanism due to collision 
between droplets that is caused by the free, thermal diffusion 
of droplets without any interaction \cite{Gunton,BS2,Siggia}.  
The former is known as Lifshitz-Slyozov-Wagner (LSW) mechanism and 
the latter as Binder-Stauffer (BS) mechanism.
Both mechanisms give the same form of the power law of 
for the evolution of the mean droplet radius $R$ 
as a function of the phase-separation time $t$, in three dimensions 
\cite{Ohta,Huse,Mazenko,Wong}: 
\begin{eqnarray}
R^3=k_d(k_{\rm B} T/5\pi \eta)t, \label{rt13}
\end{eqnarray} 
where $k_d$ is a constant, $T$ is the temperature, $\eta$ is the viscosity, 
and $k_{\rm B}$ is Boltzmann's constant. 
For LSW mechanism $k_d=0.053$, while for BS mechanism 
$k_d=6 \Phi_d$ 
($\Phi_d$: the volume fraction of the minority phase) \cite{Siggia,noteS}.  
The droplet coarsening  behavior in phase-separating  fluid mixtures 
has mostly been explained in terms of these LSW and BS mechanisms so far. 
Recently, however, some reserachers start to think that 
this well-accepted view needs some essential modifications in physics 
on the basis of experimental results and theoretical considerations. 
There are two important physical factors that have so far been 
ignored: (i) interdroplet interactions 
\cite{Kalmykov,HTIFQ2,HTJCP,HTFULL,Karpov} 
and (ii) hydrodynamic interactions \cite{HTFULL,Nikolayev}.  

In this Communication, we consider the interdroplet 
interactions between droplets and the resulting hydrodynamic motion 
of droplets \cite{noteK}.
We think that the conventional picture 
that a droplet moves around freely by Brownian motion 
and accidentally collides with another droplet 
is not correct in the exact sense, at least, for a system of a 
high droplet density \cite{HTIFQ2,HTJCP}. 

The kinetic equations for classical binary fluids 
are \cite{Gunton} 
\begin{eqnarray} 
\frac{\partial \phi}{\partial t}&=&-\vec{\nabla} \cdot (\phi \vec{v})+
L_0 \nabla^2 \frac{\delta (\beta H)}{\delta \phi}+\theta, \label{fluid1}\\
\rho \frac{\partial \vec{v}}{\partial t}&=& 
\vec{F}_\phi-\nabla p+ 
\eta\nabla^2 \vec{v}+\vec{\iota},  \label{fluid2}
\end{eqnarray}
where $\phi$ is the composition, $\vec{v}$ is the velocity, 
$\rho$ is the density, $p$ is a part of the pressure that is determined 
to satisfy the incompressibility condition, 
$\theta$ and $\vec{\iota}$ are thermal noises, and $\beta=1/k_{\rm B}T$. 
Here $H$ is the Ginzburg-Landau-type Hamiltonian: 
\begin{eqnarray}
H= k_{\rm B}T \int d \vec{r}[f(\phi)
+\frac{K}{2}(\nabla \phi)^2] \label{H}
\end{eqnarray} 
with $f(\phi)=-\frac{\tau}{2} \phi^2+\frac{u}{4} \phi^4$. 
In Eq. (\ref{fluid2}), $\vec{F}_\phi$ is the thermodynamic force density 
acting on the fluid due to the fluctuations of the composition $\phi$ 
that is given by 
\begin{eqnarray}
\frac{\vec{F}_\phi}{k_{\rm B}T}=-\phi \nabla \mu 
=-\nabla [\pi-K\nabla (\phi \nabla^2 \phi)] 
-K \nabla^2 \phi \nabla \phi, 
\nonumber
\end{eqnarray} 
where $\mu=\delta(\beta H)/\delta \phi$ is the chemical potential 
and $\pi$ is the osmotic pressure given by $\pi=\phi 
\partial f/\partial \phi -f$. 
The conserved part of the above thermodynamic force,  
$\vec{F}_\phi^{c}/k_{\rm B}T=-\nabla [\pi-K\nabla (\phi \nabla^2 \phi)]$, 
cannot produce any velocity fields under the incompressibility 
condition $\vec{\nabla}\cdot \vec{v}=0$ and it has to be 
completely balanced with a part of $\nabla p$. 
Accordingly, only a part of the other force, $\vec{F}_\phi^{int}
=-K \nabla^2 \phi \nabla \phi$, can produce the velocity 
fields through Eq. (\ref{fluid2}). 
Using the steady-state approximation 
($\partial \vec{v}/\partial t \sim 0$), thus, 
we obtain the following expression of $\vec{v}$ \cite{Gunton,Kawasaki}: 
\begin{eqnarray} 
\vec{v}&=&\int d \vec{r'} \mbox{\boldmath$T$}(\vec{r}-\vec{r'})\cdot 
\vec{F}_\phi^{int} (\vec{r'}), \label{v2}
\end{eqnarray}
where $\mbox{\boldmath$T$}$ is the so-called Oseen tensor. 

This expression gives us the physical insight on hydrodynamic effects 
on domain coarsening as follows \cite{Bray,KO,Onuki}. 
After the formation of a sharp interface, it has so far been assumed that 
the interface profile can be described by $\phi_{int}=\phi_e \tanh 
(\zeta/\sqrt 2 \xi)$ \cite{inter}, 
where $\phi_e(=\sqrt{\tau/u})$ is the equilibrium 
composition, $\xi$ is the correlation length. 
$\zeta$ is the distance from the interface defined by 
$\zeta =\vec{n} \cdot (\vec{r}-\vec{r}_{int})$, where 
$\vec{r}_{int}$ is a point on the interface and 
$\vec{n}$ is the unit normal vector at the point $\vec{r}_{int}$ 
toward the domain with a positive value of $\phi$. 
Then, the thermodynamic force density $\vec{F}_\phi^{int}$ 
can be expressed by 
\begin{eqnarray}
\frac{\vec{F}_\phi^{int}}{k_{\rm B}T}&=& 
-K (\vec{\nabla}\cdot \vec{n})(\frac{\partial \phi_{int}}{\partial \zeta})^2 
\vec{n}. \label{F2}
\end{eqnarray}
Note that $\vec{\nabla}\cdot \vec{n}$ is the curvature at $\vec{r}_{int}$ 
and $\vec{\nabla}\cdot \vec{n}=1/R_1 +1/R_2$, where 
$1/R_1$ and $1/R_2$ are the two principal curvatures of the interface. 
Here we use the relations: 
\begin{eqnarray}
\nabla \phi &\cong& (\partial \phi_{int}/\partial \zeta) \vec{n}, \label{grad} 
\\
\nabla^2 \phi &\cong& \frac{\partial^2 \phi_{int}}{\partial \zeta^2}
+(\frac{\partial \phi_{int}}{\partial \zeta}) 
\vec{\nabla}\cdot \vec{n}. \label{Laplacian}
\end{eqnarray}
Thus, we obtain 
the following equation by putting Eq. (\ref{F2}) into Eq. (\ref{fluid2}) 
and using 
$k_{\rm B}TK(\partial \phi_{int}/\partial \zeta)^2 \cong \sigma \delta(\zeta)$: 
\begin{eqnarray}
-\nabla p'-\sigma (\frac{1}{R_1}+\frac{1}{R_2}) \delta(\zeta) \vec{n} 
+\eta \nabla^2 \vec{v}=0, \label{kinetic1}
\end{eqnarray}
where $\sigma$ is the interface tension. 
It should be noted that Eq. (\ref{kinetic1}) is dependent on 
$\Phi_d$ through the characteristic curvatures of the interface.  
{\it Under the incompressibility condition, thus, 
the domain geometry ($\vec{n}$ pattern) dominates the coarsening 
mechanism}: 
For a nearly symmetric composition ($\Phi_d \sim1/2$), 
we have a bicontinuous pattern; 
and, thus, the second term in Eq. (\ref{kinetic1}) 
produces the velocity fields of $v\sim \sigma/\eta$ 
that lead to the growth law 
of $R \sim (\sigma/\eta) t$ (Siggia's mechanism). 
For an off-symmetric composition ($\Phi_d \neq 1/2$), on the other hand, 
we have a droplet pattern; and, thus, 
the second term in Eq. (\ref{kinetic1}) is balanced with $\nabla p'$ 
to satisfy $\nabla \cdot \vec{v}=0$. Thus, there is a pressure 
difference of $2\sigma/R$ across the interface (Laplace's law), 
while there are no macroscopic velocity fields $\vec v=0$ except 
during the coalescence process.  
The latter fact is the basis of Brownian-coagulation mechanism, 
which assumes that there is no interparticle interaction and 
the droplet motion is purely driven by thermal velocity noises 
[$\vec{\iota}$ in Eq. (\ref{fluid2})]. 

Here we would like to raise a fundamental question of whether 
droplets can feel neighboring droplets or not. 
The above argument tells us that the answer is ``no''. 
However, we point out that in the above argument there are  
deficiencies that can be important particularly in phase separation of 
off-symmetric binary mixture; namely, (i) 
{\it the gradual concentration change 
around a droplet is neglected above and the concentration 
profile is wrongly assumed to be always 
described by $\phi_{int}$ along the direction 
perpendicular to the interface} and (ii) {\it hydrodynamic flow induced 
by droplet collision is neglected}.  
We focus here mainly on the problem (i). 
Since the matrix can never be in equilibrium with  
droplets with different sizes at the same time, 
a droplet usually has long-range concentration (diffusion) fields around it, 
as recently pointed out by us \cite{HTIFQ2,HTJCP,HTFULL} 
and Karpov \cite{Karpov}. 
We think that this fact plays an important role in droplet coarsening, 
especially for fluid mixtures: 
The finite curvature of droplets naturally leads to the 
diffuse concentration change around them and the coupling between these 
concentration fields causes interactions between droplets. 
Experimental evidence of such interactions 
between droplets has recently been reported for fluid mixtures 
by Aver'yanov and Kalmykov \cite{Kalmykov} and Tanaka \cite{HTIFQ2}.

Before discussing the dynamics, we need to know 
the quasi-stationary concentration profile around a droplet 
for estimating the strength of the interdroplet interaction. 
After the formation of a sharp interface, 
for both spinodal decomposition (SD) and nucleation and growth (NG), 
we have $ \partial \delta \phi/\partial t =L_0 \nabla^2 \mu$ 
\cite{Mazenko,Onuki}, neglecting collision effects. 
Under the quasi-stationary approximation, thus, 
the concentration profile around a three-dimensional 
spherical droplet can be obtained by solving the Laplace equation 
$\nabla^2 \phi=0$ under the Gibbs-Thomson equation, 
\cite{Bray,LS,Karpov,Voorhees,Tokuyama} as  
$\phi=\bar\phi+(\phi_b-\bar\phi)R/r$, 
where $\bar\phi$ is the far-field concentration and 
$\phi_b$ is the concentration at the droplet boundary. 
$\phi_{b}$ is given by the following Gibbs-Thomson equation:  
$\phi_{b}=-\phi_e+(\sigma/2k_{\rm B}T \phi_e \tau)(1/R)$.  
Note that the this $1/r$ dependence of $\phi(r)$, or $\mu(r)$,  
originates from the conserved nature of the order parameter.  
For $\delta \phi$ ($=\phi-\bar{\phi}$), thus, we have the following relations:
\begin{eqnarray}
\delta \phi(r)=\delta \phi_{b} \frac{R}{r}~ (r>R), \quad
\delta \phi(r)=\delta \phi_{b}~(r<R), 
\end{eqnarray} 
where $\delta \phi_{b}=\phi_b-\bar{\phi}$.

This diffuse concentration change around droplets, 
$\delta \phi(r)$, qualitatively alters the preceding conclusion 
that there are no velocity fields for droplet phase separation 
except for thermal noises. 
In the preceding discussion, the relation 
$\nabla \phi \cong (\partial \phi_{int}/\partial \zeta) \vec{n}$ are used. 
This relation is reasonable as the first approximation and can be used 
for the case of bicontinuous phase separation, where the energy associated 
with a sharp interface governs the fluid dynamics. 
For the case of droplet phase separation, 
$\vec{F}_\phi^{int}$ given by Eq. (\ref{F2}) 
does not produce any velocity fields, 
as described before; and, thus, the first-order corrections do play 
important roles. Accordingly, we should use the 
following relation that include the first-order term of $\delta \phi$, 
instead of Eq. (\ref{grad}): 
\begin{eqnarray}
\nabla \phi &\cong& (\frac{\partial \phi_{int}}{\partial \zeta}) \vec{n}
+\nabla \delta \phi.   \label{grad2} 
\end{eqnarray} 
Here we assume that the phase having a positive value of $\phi$ 
is the minority phase. In such a case, we have the relation 
$R_1=R_2=-R$ for a spherical droplet. 
Equation (\ref{kinetic1}) should be modified 
using Eq. (\ref{grad2}) 
to include the additional thermodynamic force 
\begin{eqnarray}
\frac{\vec{F}_{grad}}{k_{\rm B}T}&=&
-K[\frac{\partial \phi_{int}}{\partial \zeta} \vec{\nabla}\cdot \vec{n}
+\nabla^2 \delta \phi]\nabla \delta \phi \nonumber \\
&\cong& \frac{K}{R}[4\phi_e+\delta \phi_{b}]\delta(\zeta) 
\nabla \delta \phi. \label{Fgrad2} 
\end{eqnarray} 
Here we use $\partial \phi_{int}/\partial \zeta \cong 2 \phi_e \delta(\zeta)$. 
We neglect the second term since it is smaller than the first term 
by a factor of $\sim \xi/R$.  
This interface force produces the following new velocity fields 
for droplet phase separation in addition to the thermal velocity noises 
$\vec{\iota}$: 
\begin{eqnarray}
\vec{v}_{grad}(\vec{r})= \int d\vec{r'} \mbox{\boldmath$T$}(\vec{r}-\vec{r'}) 
\cdot \vec{F}_{grad}(\vec{r'}). \label{vgrad} 
\end{eqnarray}
This is the theoretical basis of fluid motion induced by 
diffuse concentration change around droplets.

Next we estimate the interdroplet interaction 
{\it via} $\vec{F}_{grad}$. 
First we consider the coupling between a single spherical particle 
with a radius of $R$ at the origin and the concentration 
field of a constant gradient given by $\vec{E} \cdot \vec{r}$. 
This gradient induces the anisotropic components in the force 
given by Eq. (\ref{Fgrad2}). 
Using the analogy between this 
problem and the change of electric potential caused by a metallic 
sphere in an electric field \cite{Karpov,Voorhees}, 
we can estimate the resulting matrix concentration profile around the 
spherical droplet, $\delta \phi_E$, as 
\begin{eqnarray}
\delta \phi_E(\vec{r})=\delta \phi(\vec{r})+\vec{E}\cdot \vec{r}
-\vec{E}\cdot \vec{r} R^3/r^3. \label{phigrad}
\end{eqnarray}
Using Eq. (\ref{phigrad}), 
we obtain the total force acting on the droplet 
by integrating the terms containing $E$ in Eq. (\ref{Fgrad2}) as 
\begin{eqnarray}
\frac{\vec{F}_{drop}}{k_{\rm B}T}&=& \frac{4K\phi_e}{R} 
\int d\vec{r} \delta(|\vec{r}|-R) \nabla \delta \phi_E(\vec{r})
\nonumber \\
&=& 16 \pi K R \phi_e \vec{E}. \label{Fdrop} 
\end{eqnarray} 
Using this relation, we can straightforwardly calculate the total force 
on a droplet at $\vec{r_1}$ due to the concentration fields of a neighboring 
droplet at $\vec{r_2}$ by simply 
replacing the concentration gradient of 
$\vec{E}$ by $\nabla \delta \phi(\vec{r})$ 
($\vec{r}=\vec{r_1}-\vec{r_2}$) as  
\begin{eqnarray}
\vec{F}_{R}=-16 \pi k_BT K R^2 \phi_e \delta \phi_{b} 
\frac{\vec{r}}{r^3}.  
\label{FR}
\end{eqnarray}
Here we assume the radii of both droplets are $R$. 
This force can be easily estimated for the case of $\bar{\phi}=-\phi_e$ 
and it gives the strongest force \cite{note1}: 
$\vec{F}_{R}^{max}=-\gamma k_BT R \frac{\vec{r}}{r^3}$, 
where $\gamma =16 \pi \sigma_0 = 5 \sim 10$.  
Here, we use $\delta \phi_{b}=(2/3)\phi_e (\xi/R)$, 
$\xi^2=K/2\tau$, and $\sigma=4K \phi_e^2/3\xi \sim \sigma_0 k_{\rm B}T/\xi^2$ 
($\sigma_0=0.1 \sim 0.2$ is the universal constant). 
Here it should be noted that this force is not due to the 
interaction between two droplets in the exact sense, but due to 
the coupling between a droplet interface and the concentration gradient 
created by the neighboring droplet. 

For a fluid system, 
the resulting droplet velocity $\vec v_{R}^{max}$ can be estimated by 
balancing the frictional force $\vec F_S=-5 \pi \eta R \vec v_{R}$ and 
$\vec F_{R}^{max}$, as 
\begin{eqnarray} 
\vec v_{R}^{max} \sim \frac{\vec{F}_{R}^{max}}{5 \pi \eta R}=
\frac{\vec{F}_{R}^{max} D_{R}}{k_{\rm B}T}=-\frac{\gamma D_R R \vec{r}}{r^3}. 
\label{vR}
\end{eqnarray}
Here $D_{R}$ is the diffusion constant of a droplet 
of radius $R$ and $D_R=k_{\rm B}T/(5 \pi \eta R)$. 
It is worth noting that a simple dimensional analysis of Eq. (\ref{vR}) 
tells us that this new mechanism alone leads to the domain growth law 
of $R \sim t^{1/3}$.  

On the basis of the above results, we here estimate the maximum 
coarsening rate including both Brownian-coagulation mechanism 
(thermal noise effects) 
and the above diffusion-coupling mechanism 
(see, on the calculation technique, Refs. \cite{Siggia,Karpov,Levich}). 
Let us consider a droplet fixed at the origin in the uniform cloud 
containing $n$ droplets per volume. We suppose 
the coalescence process is much quicker than 
the collision interval, which is a reasonable approximation for a fluid 
system \cite{Siggia}. The boundary conditions for $n$ are (i) 
$n=0$ at $r=R$, $t>0$, and (ii) $n=n$ as $r \rightarrow \infty$. 
The flow of droplets to the origin $I$ is given by 
$I(r)=4 \pi r^2 j(r)$ with $j(r)=-D_R \partial n/\partial r 
+n v_{R}$.  Note that $I(r)$ should be constant with $r$. 
Using $I(r)=I$ and the above boundary conditions, we finally 
get 
\begin{eqnarray}
\frac{dn}{dt}=2nI=-\frac{8 \pi n^2 D_R R \gamma}{1-\exp(-\gamma/2)}, 
\label{et}
\end{eqnarray}
where the factor 2 accounts for the fact that the droplet at the origin is 
free to move. 
Using the relation $n=\Phi_d/(\frac{4 \pi}{3} R^3)$, 
we obtain from Eq. (\ref{et}) the following 
domain growth law: 
\begin{eqnarray}
R=[6 \gamma \Phi_d (D_R R)]^{1/3} t^{1/3}.  \label{Rt}
\end{eqnarray}
The corresponding scaled coarsening law is given by 
\begin{eqnarray}
(R/\xi)^3=6 \gamma \Phi_d \tau, 
\end{eqnarray}
where $\tau=t/\tau_\xi$ 
[$\tau_\xi=\xi^2/D_\xi$ ($D_\xi=k_BT/5 \pi \eta \xi$)].  

The time exponent of this final relation 
is the same as Brownian-coagulation mechanism, 
but the prefactor is larger by a factor of $\gamma$ than it. 
This result is quantitatively consistent with the recent careful experimental 
studies on droplet phase separation  
in space by Perrot {\it et al.} \cite{Perrot}: They 
demonstrated the $t^{1/3}$ law with the twice larger 
prefactor than that of 
Brownian-coagulation mechanism [see Eq. (\ref{rt13})], using 
the relation $k_d=12 \Phi_d$ \cite{Siggia}. 
This means $\gamma^{1/3} \sim 2.6$, if we use the relation 
$k_d=6 \Phi_d$, which we believe is the correct one \cite{noteS}. 
In our estimation $\gamma^{1/3}=1.7 \sim 2.2$ at maximum, and 
the agreement is satisfactory at least on a qualitative level. 
If we take into account the deviation 
of $\bar{\phi}$ from $\phi_e$ \cite{note1}, $\gamma^{1/3}$ 
becomes smaller to be between 1 and 2. 
We need a more analytical approach for the more accurate estimation of 
$\gamma$. 

Finally, we consider novel hydrodynamic effects unique to fluid systems 
that are not included in the above calculation: 
(i) {\bf Direct hydrodynamic interactions:} Interdroplet collision and 
the resulting shape relaxation (or the relaxation of curvature) 
directly induces 
hydrodynamic flow {\it via} Eq. (\ref{v2}) and leads directly to another 
collision. This effect likely leads to $v\sim \sigma/\eta$ and $R\sim t$ 
as first demonstrated by Nikolayev et al. \cite{Nikolayev} and also 
by us \cite{HTFULL}. 
This mechanism is likely important only for the case of a very  
high droplet density \cite{HTFULL,Nikolayev}. 
Since this problem is beyond the scope of this paper 
focusing on the growth law of $R \sim t^{1/3}$, 
we do not discuss it in detail here.  
(ii) {\bf Coupling between hydrodynamic and diffusion modes:} 
{\it The quick hydrodynamic 
coalescence after a collision could leave the droplet out of local equilibrium 
and temporally cause a strong diffusion field around it.}  
In our previous papers \cite{HTIFQ1}, 
we discuss the effects of {\it interface quench} 
induced by the quick hydrodynamic reduction of the interfacial energy, 
focusing on bicontinuous phase separation. 
We expect a similar, but qualitatively different, dynamic effect 
for the process of a quick hydrodynamic coalescence after 
interdroplet collision for droplet phase separation \cite{HTIFQ2,HTJCP}: 
The concentration profile around a droplet just after the collision 
is about the same as that before the collision 
since the diffusion process is retarded from the hydrodynamic one. 
Note that only diffusion can change the concentration profile. 
Further, the radius of the droplet increases 
from $R$ to $R'$ immediately by collision, satisfying the volume conservation. 
(a) This increases the interaction strength by a factor of $(R'/R)$ 
for a droplet experiencing 
collision and increases the probability of the subsequent collisions. 
This situation can be realized since the characteristic time of 
droplet fusion $t_{fusion} \sim \eta R/\sigma$ is generally much shorter 
than the characteristic material diffusion time for a droplet 
$t_{diff} \sim R^2/D_\xi$: 
Using the relation $\sigma \sim \sigma_0 k_{\rm B}T/\xi^2$, we obtain 
$t_{fusion}/t_{diff} \sim \xi/10 R$.  
The strong excess diffusion fields induced by collision 
lasts for about $t_{diff}$ after the interdroplet collision. 
(b) More importantly,  
the collision significantly increase the local value of $\nabla^2 \delta \phi$ 
(or $\nabla^2 \mu$) itself around the droplet for a period of $t_{diff}$ 
and, thus, produce the strong attractive interactions between the droplet 
experiencing the collision and the surrounding droplets 
{\it via} Eq. (\ref{Fgrad2}). 
This leads to the subsequent collisions.  

All these effects [(i), (ii)(a), and (ii)(b)] 
likely strengthen "gradient-induced coupling 
mechanism" selectively for droplets experiencing collision 
and lead to the "collision-induced collision" 
phenomena \cite{HTIFQ2,HTJCP} that {\it a droplet experiencing collision 
has more probability of the subsequent collisions}. 
Such behavior is actually observed experimentally 
in droplet phase separation \cite{HTIFQ2}. 
This mechanism can be important when $t_{coll} < t_{diff}$. 
Here $t_{coll}$ is the characteristic interval of droplet collision 
by the hydrodynamic translational motion of droplets. 
It is estimated 
as $t_{coll}(L)=\smallint^{L/2}_{0} dr/v_{R}^{max}=
L^3/(24D_R R \gamma)$, 
where $L$ is the average interdroplet distance. 
Since $L^3=(4 \pi/3)R^3/\Phi_d$, $t_{coll}=\pi R^2/18 \gamma \Phi_d D_R$.  
The ratio $t_{coll}/t_{diff}$ is, thus, estimated as 
$t_{coll}/t_{diff} \sim \pi R/18 \gamma\Phi_d \xi \sim 0.02 (R/\xi)\Phi_d$. 
Thus, this effect is likely important for 
the stage where $R <\sim 200 \xi$, for example, at 
$\Phi_d \sim 0.25$. This relation also suggests that 
in the very late stage the evaporation-condensation 
mechanism plays an important role. However, there have so far been no reliable 
experiments on such an extremely late stage. 

In summary, we study the coarsening mechanism of droplet phase separation. 
We find that the interdroplet interaction caused by 
the anisotropic coupling of concentration fields can  
accelerate droplet coarsening, in contrast to the 
conventional understanding that there are no interdroplet interactions, 
which is the precondition of Brownian-coagulation mechanism. 
We obtain the domain coarsening law of $R \sim t^{1/3}$, 
whose prefactor is at least larger than that of 
Brownian-coagulation mechanism.  
We call this new mechanism of droplet coarsening 
"gradient-induced coupling mechanism".  
We also demonstrate the possibility that the strong coupling between 
velocity and diffusion fields may lead to a new qualitative effect, 
"collision-induced collision". 
Further theoretical and experimental studies are highly desirable for 
the clear underastanding of this problem, including another 
effect of ``collision-induced collision via flow'' \cite{HTFULL,Nikolayev}. 

The author is grateful to T. Ohta, M. Doi, Y. Oono, T. Araki, and 
T. Kawakatsu for valuable discussions. 
He also thanks A.E. Kalmykov and V.S. Nikolayev 
for sending their papers \cite{Kalmykov,Nikolayev}. 
This work was partly supported by a 
Grant-in-Aid from the Ministry of Education, Science, and Culture, Japan, 
and also by a grant from Toray Science Foundation.

\end{multicols}

\begin{references}
\bibitem{noteS} Siggia derived $k_d=12 \Phi_d$, but we think 
the relation $k_d=6 \Phi_d$ is the right one (see also Ref. [15])
\bibitem{noteK} Karpov has recently developped  an analytical theory of 
coarsening of nuclei based on the similar idea as ours, but he has not 
considered incompressible hydrodynamics.
\bibitem{inter} Here we use the mean-field approximation for simplicity.
\bibitem{note1} It should be noted that the mean-field concentration, 
$\bar{\phi}=-\phi_e+(\sigma/2k_BT \phi_e \tau)(1/\bar{R})$ ($\bar{R}$: 
the average radius of droplets), for example, gives us slower coarsening. 
In this most realistic case, interactions between droplets are either 
attractive or repulsive, depending upon the spatial 
configuration of droplets; and, thus, we have to solve a 
more complicated many-body problem.
\end{references}
\end{document}